\documentclass[12pt]{article}
\usepackage{amssymb,amsmath,epsfig}

\begin{document}

\title{\bf Kaluza-Klein Cosmology with Varying $G$ and $\Lambda$}
\author{M. Sharif \thanks {msharif@math.pu.edu.pk} and Farida
Khanum \thanks{faridakhanum.math@yahoo.com}\\
Department of Mathematics, University of the Punjab,\\
Quaid-e-Azam Campus, Lahore-54590, Pakistan.}

\date{}

\maketitle
\begin{abstract}
In this paper we investigate non-compact FRW type Kaluza-Klein
cosmology coupled with $5D$ energy-momentum tensor. The field
equations are solved by taking gravitational and cosmological
constants as a function of time $t$. We use $\Lambda(t)=\epsilon
H^{2}$ to explore cosmological parameters including statefinder and
discuss them for dust, radiation and stiff matter dominated eras.
Also, we evaluate lookback time, proper distance, luminosity
distance and angular diameter distance. We obtain a universe which
is not compatible with current cosmological observations at
$\epsilon=6$. However, for $\epsilon>6$, the results are compatible
with observational cosmology.
\end{abstract}

\section{Introduction}

The observable universe is spatially flat and experiencing an
accelerating expansion confirmed by type Ia Supernova \cite{1},
Sloan Digital Sky Survey \cite{2}, X-ray \cite{3} and Cosmic
Microwave Background Radiations \cite{4}. It is believed that a
fluid known as dark energy (DE) with large negative pressure is
responsible for this accelerating expansion. The most suitable
representative of DE is the cosmological constant. The observational
cosmology suggests that the value of cosmological constant is
$10^{-55}cm^{-2}$ but particle physicist predicts its value
$10^{120}$ times greater than this factor. This problem is known as
the cosmological constant problem. In order to solve this problem,
some people \cite{5}-\cite{9} considered the cosmological constant
as a varying parameter.

Dirac showed that some fundamental constants do not remain constant
forever rather they vary with time due to some causal connection
between micro and macro physics \cite{10}. This is known as Large
Number Hypothesis (LNH) which gained more attention after the
discovery of accelerating expansion of the universe. In the Einstein
field equations, there involve two physical constants, i.e., the
gravitational constant $G$ (couples the geometry and matter) and
cosmological constant $\Lambda$ (vacuum energy in space). According
to the LNH, the gravitational constant should also vary with time.
Zeldovich \cite{11} extended LNH by taking cosmological constant as
$\Lambda=\frac{8\pi G^{2}m^{2}_{p}}{h^{4}}$, where $m^{2}_{p}$ is
the mass of proton and $h$ is the Planck's constant. He showed that
$\Lambda$ produces the same gravitational effects in vacuum as that
produced by matter. Consequently, this cosmological term must be
included in the physical part of the field equations. He also
defined gravitational energy of the vacuum as the interactions of
virtual particles separated by a distance $\frac{h}{m_{p} c}$, where
$c$ is the speed of light.

Many variants of $\Lambda$ have been proposed for the better
understanding of DE \cite{12}. Ray and Mukhopadhay \cite{13} solved
the field equations by using three different forms of the
cosmological constant, i.e., $\Lambda \sim(\frac{\dot{a}}{a})^{2},~
\Lambda\sim(\frac{\ddot{a}}{a})$, and $\Lambda\sim\rho$. It is shown
that these models yield equivalent results to the FRW spacetime.
Thus an investigation about the scale factor and other cosmological
parameters with varying G and $\Lambda$ may be interesting.

Over the years, addition of dimensions and unification of forces
have been irresistible for the researchers. Minkowski showed that
time can be interpreted as a fourth part of special relativity.
Similarly, electricity and magnetism were unified by Maxwell. In
this connection, next step was to unify General Relativity (GR) and
electromagnetism. The theory of Kaluza-Klein (KK)
\cite{14}-\cite{15} is the result of such an attempt. It is a $5$
dimensional GR in which extra dimension is used to couple the
gravity and electromagnetism. Overuin and Wesson \cite{16}-\cite{18}
presented an excellent review of this theory. In a recent paper,
Darabi \cite{19} studied the non-compact, non-Ricci KK theory and
coupled the flat universe with non-vacuum states of the scalar
field. He showed that the reduced field equations with suitable
equation of state (EoS) describe the early inflation and late time
acceleration.

Shani et al. \cite{20} introduced a new cosmological diagnostic pair
$(r,s)$ called statefinder which allows one to explore the
properties of DE independent of model. This pair depends on the
third derivative of the scale factor, $a$, just like the dependence
of Hubble and deceleration parameter on first and second derivative
of respectively. It is used to distinguish flat models of the DE. In
recent literature \cite{21}-\cite{29}, this pair has been evaluated
for different models. Jamil and Debnath \cite{29} solved the field
equations of the flat FRW universe with variable $G$ and $\Lambda$.
Pradhan \cite{29a} discussed solutions of the field equations and
their consequences for the flat KK universe with variable $\Lambda$
but keeping $G$ fixed.

In this paper, we take non-Ricci, non-compact FRW type KK cosmology
with variables $G$ and $\Lambda$ and solve the field equations. We
evaluate different cosmological parameters with the assumption that
our universe is filled with different types of matter. The scheme of
the paper is as follows. In the next section, the KK model and its
field equations are presented. In section \textbf{3}, solution of
the field equations and discussion on the cosmological parameters is
given. Section \textbf{4} is devoted to compute lookback time,
proper distance, luminosity distance and angular diameter distance.
In section \textbf{5}, we summarize the results.

\section{Model and the Field Equations}

The geometry of the Kaluza-Klein cosmology is defined by the
following metric \cite{29a}
\begin{equation}\label{1}
ds^{2}=dt^{2}-a^{2}(t)[\frac{dr^{2}}{1-kr^{2}}+r^{2}(d\theta^{2}
+\sin\theta^{2}d\phi^{2})+(1-kr^{2})d\psi^{2}],
\end{equation}
where $a(t)$ is the scale factor, $k=-1,0,1$ is the curvature
parameter for spatially closed, flat and open universe respectively.
We assume that the universe is filled with perfect fluid whose
energy-momentum tensor is given by
\begin{equation}\label{2}
T_{\mu\nu}=(p+\rho)u_{\mu}u_{\nu}-pg_{\mu\nu},\quad(\mu,\nu=0,1,2,3,4),
\end{equation}
where $u_{\mu}$ is the five velocity satisfying
$u^{\mu}u_{\mu}=1$, $\rho$ is the energy density and $p$ is the
pressure of cosmic fluid. The Einstein field equations are given
by
\begin{equation}\label{3}
R_{\mu\nu}-\frac{1}{2}Rg_{\mu\nu}+\Lambda g_{\mu\nu}=8\pi
GT_{\mu\nu},
\end{equation}
where $R_{\mu\nu},~g_{\mu\nu}$ and $R$ are the Ricci tensor, metric
tensor and Ricci scalar respectively. Here we take the cosmological
constant and gravitational constant, respectively, as
$\Lambda=\Lambda(t),~G=G(t)$. Using Eqs.(\ref{1})-(\ref{3}), it
follows that
\begin{eqnarray}\label{4}
8\pi
G\rho+\Lambda&=&6\frac{\dot{a}^{2}}{a^{2}}+6\frac{k}{a^{2}},\\\label{5}
8\pi
Gp-\Lambda&=&-3\frac{\ddot{a}}{a}-3\frac{\dot{a}^{2}}{a^{2}}-3\frac{k}{a^{2}}.
\end{eqnarray}

For a flat universe $k=0$, these equations reduce to
\begin{eqnarray}\label{4a}
8\pi G\rho+\Lambda=6\frac{\dot{a}^{2}}{a^{2}}=6H^{2},\\\label{5a}
8\pi Gp-\Lambda=-3\frac{\ddot{a}}{a}-3\frac{\dot{a}^{2}}{a^{2}},
\end{eqnarray}
where $H=\frac{\dot{a}}{a}$ is the Hubble parameter. The continuity
equation, $T^{\mu\nu};_{\nu}=0$, yields
\begin{equation}\label{6}
\dot{\rho}+4H(\rho+p)=0
\end{equation}
while the EoS is $p=(\omega-1)\rho,\quad\omega\in[1,2]$. Using
Eqs.(\ref{4a}) and (\ref{5a}), we have
\begin{equation}\label{8}
8\pi G[\dot{\rho}+4H(\rho+p)]+8 \pi \dot{G}\rho+\dot{\Lambda}=0.
\end{equation}
Substituting Eq.(\ref{6}) in this equation, it follows that
\begin{equation}\label{9}
8\pi\dot{G}\rho+\dot{\Lambda}=0.
\end{equation}
This shows that $\Lambda$ is constant whenever $G$ is constant and
vice versa. Using EoS in (\ref{6}) and solving, we obtain
\begin{equation}\label{10}
\rho=\frac{c_{1}}{a^{4\omega}},
\end{equation}
where $c_{1}$ is an integration constant which can be found by using
$\omega=\omega_{0}$ and $\rho=\rho_{c}$ (critical density) at time
$t=t_{0}$, the subscript $0$ denotes the present value. Thus
Eq.(\ref{10}) becomes
\begin{equation}\label{11}
\rho=\frac{\rho_{c} a_{0}^{4\omega_{0}}}{a^{4\omega}}.
\end{equation}
The deceleration parameter $q$ is defined by
\begin{equation}\label{12}
q=-1-\frac{\dot{H}}{H^{2}},
\end{equation}
where $q\leq-1$ for an accelerating universe. Inserting this value
of $q$ in Eqs.(\ref{4a}) and (\ref{5a}), we obtain
\begin{eqnarray}\label{13}
8\pi Gp&=&3H^{2}(q-1)+\Lambda,\\\label{14} 8\pi
G\rho&=&6H^{2}-\Lambda.
\end{eqnarray}

\section{Solution of the Field Equations }

In this section, we solve the field equations and evaluate some
cosmological parameters. We see that Eqs.(\ref{4a}) and (\ref{5a})
indicate the dependence of $\Lambda$ on
$G\rho,~(\frac{\dot{a}}{a})^2,~\frac{\ddot{a}}{a}$ and $a^{-2}$.
However, the case $\Lambda\sim a^{-2}$ is not consistent for $k=0$.
Here we assume the following value of $\Lambda$
\begin{equation}\label{15}
\Lambda(t)=\epsilon H^{2},
\end{equation}
where $\epsilon$ is an arbitrary constant. From Eqs.(\ref{4a}),
(\ref{5a}) and (\ref{15}) with EoS, it follows that
\begin{equation*}
\dot{H}+\omega(6-\epsilon)=0
\end{equation*}
which yields
\begin{equation}\label{17}
H(t)=-\frac{3}{(\epsilon-6)\omega t+c_{2}}.
\end{equation}
where $c_{2}$ is another constant of integration and can be
determined by using $\omega=\omega_{0},~H=H_{0}$ at $t=t_{0}$, i.e.,
\begin{equation}\label{18}
c_{2}=-\frac{3}{H_{0}}-\omega_{0}t_{0}(6-\epsilon).
\end{equation}
Thus we obtain
\begin{equation}\label{19}
H(t)=\frac{\dot{a}}{a}=\frac{3H_{0}}{3+H_{0}(6-\epsilon)(\omega
t-\omega_{0}t_{0})}
\end{equation}
which leads to
\begin{equation}\label{20}
a(t)=c_{3} [3+H_{0}(6-\epsilon)(\omega
t-\omega_{0}t_{0})]^{\frac{3}{\omega(6-\epsilon)}},
\end{equation}
where $c_{3}$ is the constant of integration. Substituting this
value of $a$ in Eq.(\ref{11}), we have
\begin{equation}\label{21}
\rho(t)=\rho_{c}a_{0}^{4\omega_{0}}c^{-4\omega}_{3}
[3+H_{0}(6-\epsilon)(\omega
t-\omega_{0}t_{0})]^{\frac{12}{\epsilon-6}}.
\end{equation}
Using Eq.(\ref{19}) in (\ref{15}), we have
\begin{equation}\label{22}
\Lambda(t)=\frac{9\epsilon H_{0}^2}{[3+H_{0}(6-\epsilon)(\omega
t-\omega_{0}t_{0})]^{2}}.
\end{equation}
Substituting Eq.(\ref{21}), the derivative of (\ref{22}) in
(\ref{9}) and integrating, it follows that
\begin{equation}\label{23}
G(t)=\frac{H^{2}_{0}(6-\epsilon)[3+H_{0}(6-\epsilon)(\omega
t-\omega_{0}t_{0})]^{\frac{2\epsilon}{6-\epsilon}}}{4\pi\rho_{c}
a^{4\omega_{0}}_{0}c^{-4\omega}_{3}}.
\end{equation}
Inserting the value of $H$ from Eq.(\ref{19}) in (\ref{12}), the
deceleration parameter becomes
\begin{equation}\label{24}
q=-1+\frac{\omega(6-\epsilon)}{3}.
\end{equation}
We take $\epsilon\geq6$ so that $q\leq-1$ for the accelerating
universe.

For $(4+1)D$ spacetime, the expansion scalar can be defined as
\cite{19a}
\begin{equation}\label{25}
\Theta=4H=\frac{12H_{0}}{H_{0}(6-\epsilon)(\omega
t-\omega_{0}t_{0})+3}.
\end{equation}
The statefinder pair $(r,s)$ is defined as follows \cite{20}
\begin{eqnarray}\label{26}
r=\frac{\dddot{a}}{aH^{3}},\quad s=\frac{r-1}{3(q-\frac{1}{2})}.
\end{eqnarray}
Using Eqs.(\ref{19}), (\ref{20}) and (\ref{24}) in (\ref{26}), we
have
\begin{equation}\label{27}
r=\frac{[3+\omega(\epsilon-6)][3+2\omega(\epsilon-6)]}{9},\quad
s=\frac{2\omega(6-\epsilon)}{9}.
\end{equation}
Notice that if we take $\epsilon=6$, Eq.(\ref{4a}) yields $8 \pi G
\rho=0$ which implies that either $G=0$ or $\rho=0$. The case $G=0$
yields $\Lambda$ to be constant (\ref{9}), which contradicts our
assumption of varying $\Lambda$. For $\rho=0$, EoS gives $p=0$ and
hence Eq.(\ref{5a}) yields $\dot{H}=0$ implying that $H(t)=constant$
for all time which is not possible. Thus $\epsilon=6$ is not
compatible with current cosmological observations.

For $\epsilon>6$, solution of the field equations yields the
following physical and geometrical parameters
\begin{eqnarray}\nonumber
a(t)&=&c_{3}[3+H_{0}(6-\epsilon)(\omega
t-\omega_{0}t_{0})]^{\frac{3}{\omega(6-\epsilon)}},\\\nonumber
\rho(t)&=&\rho_{c}a_{0}^{4\omega_{0}}c^{-4\omega}_{3}
[3+H_{0}(6-\epsilon)(\omega
t-\omega_{0}t_{0})]^{\frac{12}{\epsilon-6}},\\\nonumber
\Lambda(t)&=&\frac{9\epsilon H_{0}}{[3+H_{0}(6-\epsilon)(\omega
t-\omega_{0}t_{0})]^{2}}, \\\nonumber
G(t)&=&\frac{H^{2}_{0}(6-\epsilon)[3+H_{0}(6-\epsilon)(\omega
t-\omega_{0}t_{0})]^{\frac{2\epsilon}{6-\epsilon}}}{4\pi \rho_{c}
a^{4\omega_{0}}_{0}c^{-4\omega}_{3}},\\\nonumber
q&=&-1+\frac{\omega(6-\epsilon)}{3},\\\nonumber
\Theta&=&\frac{12\epsilon H_{0}}{H_{0}(6-\epsilon)(\omega
t-\omega_{0}t_{0})+3},\\\nonumber
r&=&\frac{[3+\omega(\epsilon-6)][3+2\omega(\epsilon-6)]}{9},\\\label{A}
s&=&\frac{2\omega(6-\epsilon)}{9}.
\end{eqnarray}
Now we evaluate the above parameters for dust, radiation and stiff
matter dominated eras by substituting $\omega=1,~\omega=4/3$ and
$\omega=2$ respectively as follows.

\subsubsection*{Case 1 ($\omega=1$):}

In this case, we assume that our universe is five dimensional RW
type non-vacuum, non-compact KK cosmology filled with dust matter.
This case includes the baryonic and non-baryonic matter and is
called the dust dominated universe. Substituting $\omega=1$ in
Eqs.(\ref{A}) for $\epsilon>6$, we have
\begin{eqnarray}\nonumber
a(t)&=&c_{3} [3+H_{0}(6-\epsilon)(
t-\omega_{0}t_{0})]^{\frac{3}{(6-\epsilon)}},\nonumber\\
\rho(t)&=&\rho_{c} a_{0}^{4\omega_{0}}c^{-4}_{3}
[3+H_{0}(6-\epsilon)( t-\omega_{0}t_{0})]^{\frac{12}{\epsilon-6}},\nonumber\\
\Lambda(t)&=&\frac{9\epsilon H_{0}}{[3+H_{0}(6-\epsilon)(
t-\omega_{0}t_{0})]^{2}},\nonumber\\
G(t)&=&\frac{H^{2}_{0}(6-\epsilon)[3+H_{0}(6-\epsilon)(
t-\omega_{0}t_{0})]^{\frac{2\epsilon}{6-\epsilon}}}{4\pi \rho_{c}
a^{4\omega_{0}}_{0}c^{-4}_{3}}, \nonumber
\end{eqnarray}
\begin{eqnarray}
q&=&-1+\frac{(6-\epsilon)}{3},\nonumber\\
\Theta&=&\frac{12\epsilon H_{0}}{H_{0}(6-\epsilon)
(t-\omega_{0}t_{0})+3},\nonumber\\
r&=&\frac{[3+(\epsilon-6)][3+2(\epsilon-6)]}{9},\nonumber\\
s&=&\frac{2(6-\epsilon)}{9}.
\end{eqnarray}
We see that all the parameters depend on $\epsilon$ and time $t$ if
$a_{0}$ and $H_{0}$ are known. Notice that for
$t\longrightarrow\infty$, the scale factor, energy density and
gravitational constant diverge while the cosmological constant
vanishes. Also, the expansion scalar becomes zero indicating no
expansion in the universe at later times.

\subsubsection*{Case 2 ($\omega=4/3$):}

Here we assume that the universe is filled with radiations. In this
case, Eqs.(\ref{A}) become
\begin{eqnarray}\nonumber
a(t)&=&c_{3}[3+H_{0}(6-\epsilon)(\frac{4}{3}
t-\omega_{0}t_{0})]^{\frac{9}{4(6-\epsilon)}},\nonumber\\
\rho(t)&=&\rho_{c}a_{0}^{4\omega_{0}}c^{\frac{16}{3}}_{3}
[3+H_{0}(6-\epsilon)(\frac{4}{3}t-\omega_{0}t_{0})]^{\frac{12}{\epsilon-6}},\nonumber\\
\Lambda(t)&=&\frac{9\epsilon H_{0}}{[3+H_{0}(6-\epsilon)(\frac{4}{3}
t-\omega_{0}t_{0})]^{2}},\nonumber\\
G(t)&=&\frac{H^{2}_{0}(6-\epsilon)[3+H_{0}(6-\epsilon)(\frac{4}{3}
t-\omega_{0}t_{0})]^{\frac{2\epsilon}{6-\epsilon}}}{4\pi \rho_{c}
a^{4\omega_{0}}_{0}c^{\frac{-16}{3}}_{3}},\nonumber\\
q&=&-1+\frac{4(6-\epsilon)}{9},\nonumber\\
\Theta&=&\frac{36\epsilon H_{0}}{H_{0}(6-\epsilon)(4
t-\omega_{0}t_{0})+9},\nonumber\\
r&=&\frac{[9+4(\epsilon-6)][9+8(\epsilon-6)]}{27},\nonumber\\
s&=&\frac{8(6-\epsilon)}{27}.
\end{eqnarray}
It is obvious from the above results that all the parameters depend
on $\epsilon$ and time $t$.

\subsubsection*{Case 3 ($\omega=2$):}

If the universe is supposed to be filled with stiff matter then the
corresponding parameters take the form
\begin{eqnarray}\nonumber
a(t)&=&c_{3}[3+H_{0}(6-\epsilon)(2
t-\omega_{0}t_{0})]^{\frac{3}{2(6-\epsilon)}},\\\nonumber
\rho(t)&=&\rho_{c} a_{0}c^{-4\omega_{0}}_{3} [3+H_{0}(6-\epsilon)(2
t-\omega_{0}t_{0})]^{\frac{12}{\epsilon-6}},
\end{eqnarray}
\begin{eqnarray}\nonumber
\Lambda(t)&=&\frac{9\epsilon H_{0}}{[3+H_{0}(6-\epsilon)(2
t-\omega_{0}t_{0})]^{2}},\\\nonumber
G(t)&=&\frac{H^{2}_{0}(6-\epsilon)[3+H_{0}(6-\epsilon)(2
t-\omega_{0}t_{0})]^{\frac{2\epsilon}{6-\epsilon}}}{4\pi \rho_{c}
a^{4\omega_{0}}_{0}c^{-8}_{3}}, \\\nonumber
q&=&-1+\frac{2(6-\epsilon)}{3},\\\nonumber \Theta&=&\frac{12\epsilon
H_{0}}{H_{0}(6-\epsilon)(2 t-\omega_{0}t_{0})+3},\\ \nonumber
r&=&\frac{[3+2(\epsilon-6)][3+4(\epsilon-6)]}{9},\nonumber\\
s&=&\frac{4(6-\epsilon)}{9}.
\end{eqnarray}
It is mentioned here that cases of radiation and stiff matter show
the same behavior as that of the dust case for late time.

\section{Observable Distances}

Int this section, we discuss some of the consequences of our work.
In the expanding universe there are many ways to define the distance
between two co-moving objects because the distance between them
changes constantly. In the following, we evaluate lookback time and
some types of distance like proper, luminosity and angular diameter
distances \cite{20a}.

\subsection{Lookback Time}

The lookback time is actually the total time travelled by a photon.
If a photon is emitted by a source at time $t_{0}$ and is received
by an observer at time $t$ then the look back time $t-t_{0}$ is
defined as
\begin{equation}\label{28}
t-t_{0}=\int^{a}_{a_{0}}\frac{da}{\dot{a}},
\end{equation}
where $a_{0}$ is the present value of the scale factor. Substituting
$\omega=\omega_{0}$ at $t=t_{0}$ in Eq.(\ref{20}), we have
\begin{equation}\label{29}
a_{0}=c_{3}3^{\frac{3}{\omega_{0}(6-\epsilon)}}.
\end{equation}
The redshift variable $z$ in terms of scale factor is defined as
\begin{equation}\label{30}
1+z=\frac{a_{0}}{a}.
\end{equation}
Equation (\ref{19}) yields the following expression for time $t$
\begin{equation}\label{31}
t=\frac{2}{\omega
H_{0}(\epsilon-6)}\left(1-\frac{H_{0}}{H}\right)+\frac{\omega_{0}t_{0}}{\omega}
\end{equation}
which implies that
\begin{equation}\label{32}
t-t_{0}=\frac{2}{\omega
H_{0}(\epsilon-6)}\left(1-\frac{H_{0}}{H}\right)+\left(\frac{\omega_{0}}{\omega}-1\right)t_{0}.
\end{equation}
Using Eqs.(\ref{20}), (\ref{29}) and (\ref{30}) in (\ref{32}), it
follows that
\begin{equation}\label{33}
t-t_{0}=\frac{2}{H_{0}\omega(\epsilon-6)}+\left(\frac{\omega_{0}}{\omega}-1\right)t_{0}
+\frac{3^{\frac{\omega_{0}}{\omega}}(1+z)^{\frac{-\omega(6-\epsilon)}{3}}}
{\omega(6-\epsilon)H_{0}}
\end{equation}
which is the lookback time. When $z\longrightarrow\infty$, the value
of lookback time or radiation traveled time for the early universe
becomes
\begin{equation}\label{33a}
t-t_{0}\sim\frac{2}{H_{0}\omega(\epsilon-6)}+\left(\frac{\omega_{0}}{\omega}-1\right)t_{0}
\end{equation}
and for late times, i.e., $z\longrightarrow-1$, it turns out to be
\begin{equation}\label{33b}
t-t_{0}\sim \frac{3^{\frac{\omega_{0}}{\omega}}(1+z)^{\frac{-\omega(6-\epsilon)}{3}}}
{\omega(6-\epsilon)H_{0}}.
\end{equation}

\subsection{Proper Distance}

The proper distance between a source and observer is given by
\begin{equation}\label{34}
d=\int^{a_{0}}_{a}\frac{da}{a\dot{a}}=a_{0}\int^{t_{0}}_{t}\frac{dt}{a}.
\end{equation}
Using Eqs.(\ref{20}), (\ref{30}) and (\ref{31}), this turns out to
be
\begin{equation}\label{35}
d=\frac{3^{\frac{\omega}{\omega_{0}}}(1+z)^{1+\frac{\omega(\epsilon-6)}{3}}}
{H_{0}\left(3+\omega(\epsilon-6)\right)}
-\frac{3}{H_{0}(3+\omega_{0}(\epsilon-6))}.
\end{equation}
When $\epsilon=6$, it reduces to
\begin{equation}\label{35a}
d=\frac{3^{\frac{\omega}{\omega_{0}}-1}}{H_{0}}-\frac{1}{H_{0}}.
\end{equation}
For the early universe, $d\longrightarrow\infty$ while for late time
universe, we have
\begin{equation}\label{35b}
d\sim-\frac{3}{H_{0}(3+\omega_{0}(\epsilon-6))}.
\end{equation}

\subsection{Luminosity Distance}

The luminosity distance is defined as
\begin{equation}\label{36}
d_{L}=\left(\frac{L}{4\pi l}\right)^{\frac{1}{2}},
\end{equation}
where $L$ is the total energy emitted by a source per unit time
while $l$ is the apparent luminosity of the object. Luminosity
distance is an important concept of theoretical cosmology. This is a
way of expanding the light coming from a distant object. It is not
the actual distance because inverse square law does not hold in real
universe. To calculate the luminosity distance the inverse square
law of brightness is generalized  from static Euclidean space to an
expanding curved space by the following expression \cite{31}
\begin{equation}\label{37}
d_{L}=d(1+z).
\end{equation}
Substituting the value of $d$ from Eq.(\ref{35}), it follows that
\begin{equation}\label{38}
d_{L}=\frac{3^{\frac{\omega}{\omega_{0}}}(1+z)^{2+\frac{\omega(\epsilon-6)}
{3}}}{H_{0}(3+\omega(\epsilon-6))}
-\frac{3(1+z)}{H_{0}(3+\omega_{0}(\epsilon-6))}.
\end{equation}
This is the required expression for luminosity distance. In this
model, the luminosity distance depends on the value of $\epsilon$
and $\omega$. If $\epsilon=6$, we obtain
\begin{equation}\label{38a}
d_{L}=3^{\frac{\omega}{\omega_{0}}-1}(1+z)^{2}-(1+z).
\end{equation}
For early inflationary epoch, i.e., $z\longrightarrow\infty$, this
does not remain finite while for late time universe, i.e.,
$z\longrightarrow-1$, this tends to zero.

\subsection{Angular Diameter Distance}

The angular diameter of a light source of proper distance $D$
observed at time $t=t_{0}$ is defined by
\begin{equation}\label{39}
\delta=\frac{D(1+z)^{2}}{d_{L}},
\end{equation}
where the angular diameter distance $d_{A}$ is defined as the
ratio of the source diameter to its angular diameter, i.e.,
\begin{equation}\label{40}
d_{A}=\frac{D}{\delta}=d_{L}(1+z)^{-2}
\end{equation}
which can also be written as
\begin{equation}\label{40a}
d_{A}=d_{L}(1+z)^{-2}=d(1+z)^{-1}=\frac{d^2}{d_L}.
\end{equation}
This indicates the relationship between angular diameter distance
and luminosity distance. Using Eq.(\ref{38}) in (\ref{40}), we have
\begin{equation}\label{41}
d_{A}=\frac{3^{\frac{\omega}{\omega_{0}}}(1+z)^{\frac{\omega(\epsilon-6)}{3}}}
{H_{0}(3+\omega(\epsilon-6))}
-\frac{3(1+z)^{-1}}{H_{0}(3+\omega_{0})(\epsilon-6)}.
\end{equation}
Notice that the value of angular diameter distance is maximum at
$z_{max}$ given by
\begin{equation}\label{42}
z_{max}=\left[\frac{3^{2-\frac{\omega}{\omega_{0}}}(2+\omega(\epsilon-6))}
{\omega(6-\epsilon)(2+\omega_{0}(\epsilon-6))}\right]^{\frac{3}{3+\omega(\epsilon-6)}}-1.
\end{equation}

\section{Conclusion}

This paper is devoted to solve the field equations for KK cosmology
by varying $G$ and $\Lambda$ with respect to time. For the case
$\Lambda(t)=\epsilon H^{2}$, we have solved the field equations and
also determined the values of some geometrical and physical
parameters. It is shown that all the parameters depend upon
$\epsilon,~t$ and $\omega$. In particular, we have explored the two
cases, i.e., $\epsilon=6$ and $\epsilon>6$.

For $\epsilon=6$, we have obtained the following results:
\begin{itemize}
\item  Here $H=constant$ for all time which is not possible.
\item The deceleration parameter is equal to $-1$ being
$\dot{H}=0$.
\item The energy density and pressure become zero.
\item The statefinder pair $(r,s)$ turns out to be $(0,\frac{2}{9})$.
\item The expansion scalar vanishes indicating that there is no
expansion.
\item The lookback time is divergent.
\end{itemize}

The case $\epsilon>6$ provides consistent results with current
cosmological observations. In this case, we have evaluated all
the physical and geometrical parameters for dust, radiation and
stiff matter dominated eras. It is found that in these eras
$a(t),~\rho(t)$ and $G(t)$ diverge when $t\rightarrow\infty$ whereas
$\Lambda(t)$ and $\Theta$ vanish. This shows that the KK universe
becomes static at later time.

We have also investigated lookback time and observable distances,
i.e., proper distance, luminosity distance and angular diameter
distance. It is found that all these distances depend upon
$\epsilon,~t,~\omega$ and the redshift variable. The maximum angular
diameter distance is also indicated.

\end{document}